\documentclass[pra,english,preprintnumbers,amsmath,amssymb,nofootinbib,twocolumn,superscriptaddress]{revtex4}
\pdfoutput=1
\usepackage[latin1]{inputenc}
\usepackage{graphicx}
\usepackage{color}
\usepackage{bbm}
\usepackage{amssymb}
\usepackage{amsmath}
\usepackage{tabularx}
\usepackage{physics}
\usepackage{dsfont}
\usepackage{multirow}
\usepackage{dcolumn}
\usepackage{booktabs}
\usepackage{braket}

\def\0#1#2{\frac{#1}{#2}}

\def\s0#1#2{\mbox{\small{$ \frac{#1}{#2} $}}}

\newcommand{\beq}{\begin{equation}}
\newcommand{\eeq}{\end{equation}}
\newcommand{\bea}{\begin{eqnarray}}
\newcommand{\eea}{\end{eqnarray}}

\usepackage{babel}
\makeatother

\begin{document}

\title{Thermodynamics and static response of anomalous 1D fermions \\ via quantum Monte Carlo in the worldline representation}

\author{J. R. McKenney}
\affiliation{Department of Physics and Astronomy, University of North Carolina. Chapel Hill, North Carolina 27599, USA}
\author{A. Jose}
\affiliation{Department of Physics and Astronomy, University of North Carolina. Chapel Hill, North Carolina 27599, USA}
\author{J. E. Drut}
\affiliation{Department of Physics and Astronomy, University of North Carolina. Chapel Hill, North Carolina 27599, USA}

\begin{abstract}
A system of three-species fermions in one spatial dimension (1D) with a contact three-body interaction is known to display a scale anomaly. 
This anomaly is identical to that of a two-dimensional (2D) system of two-species fermions. The exact relation between those two systems, however, is limited to the two-particle sector of the 2D case and the three-particle sector of the 1D case. 
Here, we implement a non-perturbative Monte Carlo approach, based on the worldline representation, to calculate the thermodynamics and static response of three-species fermions in 1D, thus tackling the many-body sector of the problem. 
We determine the energy, density, and pressure equations of state, and the compressibility and magnetic susceptibility for a wide
range of temperatures and coupling strengths. We compare our results with the third-order virial expansion.
\end{abstract}

\date{\today}

\maketitle 

\section{Introduction}

The few- and many-body physics of one-dimensional (1D) fermionic systems has been a subject of
active study for decades. The availability of exact solutions (namely the Bethe ansatz~\cite{BetheAnsatzReview}), and powerful numerical 
(e.g. density-matrix renormalization group~\cite{RevModPhys.77.259}) as well as analytic (such as bosonization, see e.g.~\cite{von_Delft_1998}) approaches to study these problems
has elucidated their properties in a way that has not been matched in higher dimensions. That being said,
by far most of the work referred to above concerns systems with only two-body interactions.

Recently, a series of papers~\cite{PhysRevA.97.011602, PhysRevA.97.061605, PhysRevA.97.061604, PhysRevLett.120.243002, PhysRevA.97.061603} 
showed that fine-tuned 1D systems (bosonic or fermionic) with {\it only three-body interactions} display the remarkable property of anomalous scale invariance. 
These systems are classically invariant under scale transformations as they possess no intrinsic scale, but their quantum mechanical version requires the introduction 
of a new scale (namely a regulator) and furthermore bound states appear in the case of attractive interactions. The appearance of such a new scale, often called
dimensional transmutation, is well understood and has been described by many authors and in a variety of situations (see e.g.~\cite{JackiwDelta,holstein,PhysRevA.55.R853, PhysRevLett.109.135301, hoffman, PhysRevA.88.043636, ORDONEZ201664, The2dPaper}). In particular, Ref.~\cite{PhysRevLett.120.243002} found that the 
1D three-body dynamics with a three-body interaction is identical to that of two particles in 2D with a two-body interaction, and from there derived cross-dimensional 
connection between virial coefficients.

The references pointed out above, and a few that followed~\cite{PhysRevA.99.013615, PhysRevA.99.053607, PhysRevA.100.063604}, began exploring several 
properties of these remarkable systems. In this work we push those investigations forward by carrying out a fully nonperturbative calculation 
of the thermodynamics (including static response) of three-component fermions in 1D with attractive interactions. 
Using a lattice formulation and the worm algorithm form of quantum Monte Carlo, we explore a wide range of coupling and 
temperature regimes, furnishing predictions for future experiments.

The formulation and algorithm used here follow closely those of Ref.~\cite{PhysRevD.99.074511} (see also Ref.~\cite{PhysRevA.85.063624}) and can be 
easily generalized to include two-body as well as four-body forces and beyond (such as those originally proposed in Ref.~\cite{PhysRevA.82.043606} and studied in detail in Refs.~\cite{PhysRevLett.109.250403, PhysRevA.87.063617}), 
without suffering from a sign problem (in 1D). This indicates that a vast number of systems are open for research with the methods used here. 
Furthermore, the experimental realization of multi-flavor systems with many-body forces is currently under intense investigation~\cite{goban}.

\section{Hamiltonian, lattice theory, and many-body formalism}

We will study a system of three-species fermions in 1D governed by the following Hamiltonian:
\beq
\label{Eq:H}
\hat H = \hat T + \hat V,
\eeq
where the kinetic energy operator is $\hat T = \hat T_1 + \hat T_2 + \hat T_3$,
where
\beq
\hat T_s = \int dp \; \frac{p^2}{2m}\hat a_{s}^\dagger(p) \hat a_{s}(p),
\eeq
and the interaction energy operator is
\beq
\hat V = g \int dx \; \hat n_1(x)  \hat n_2(x)  \hat n_3(x).
\eeq
Here, $\hat a_{s}^\dagger(p)$ and $\hat a_{s}(p)$ are the fermionic creation and annihilation operators
for particles of species $s$ and momentum $p$, and $\hat n_s(x)$ is the corresponding density
at position $x$. From this point on, we will take $m = \hbar = k_B = 1$.

To calculate the thermodynamics of this system, in particular the
static response in the spin and density channels, we discretize spacetime and
use a nonperturbative Monte Carlo method, as explained next.
Our starting point is the grand-canonical partition function
\beq\label{Eq:Z}
\mathcal Z = \text{Tr}\left [ e^{-\beta(\hat H -\mu \hat N)} \right],
\eeq
where the trace is over the Fock space, $\hat H$ is as in Eq.~(\ref{Eq:H}),  $\hat N = \hat N_1 + \hat N_2 + \hat N_3$ is the 
total particle number operator, and $\hat N_i$ is the particle number operator for the $i$-th species. We will focus on unpolarized 
systems in this work, such that the chemical potential $\mu$ is the same for all species.

To address the Boltzmann operator when both $\hat T$ and $\hat V$ are present, 
taking into account that those operators do not commute, we implement the following 
steps. First, we will place the system in a spatial lattice and choose a nearest-neighbor 
discretization of the kinetic energy, namely
\beq
\hat T_s = -\frac{1}{2}\sum_{i,j} \hat \psi^\dagger_{s,i} (\delta_{i,j+1} + \delta_{i,j-1} - 2 \delta_{i,j}) \hat \psi^{}_{s,j},
\eeq
where $\hat \psi^{\dagger}_{s,i}$ and $\hat \psi^{}_{s,i}$ respectively create and annihilate a fermion 
of species $s$ at location $i$ on the lattice. 

Second, we write the on-site interaction as
\beq
\hat V = g \sum_i \; \hat n_{1,i} \hat n_{2,i}  \hat n_{3,i},
\eeq
where $\hat n_{s,i} = \hat \psi^{\dagger}_{s,i} \hat \psi^{}_{s,i}$, and the total particle number is therefore
$\hat N = \hat N_1 + \hat N_2 + \hat N_3$, where
\beq
\hat N_s = \sum_{i} \hat n_{s,i}.
\eeq

Using the above discretized forms (note that we have taken the spatial lattice spacing to be $\ell = 1$, but we will retain $\tau$ 
as the imaginary-time lattice spacing), we may separate $\hat H -\mu \hat N$ into diagonal (or ``on-site'') and off-diagonal pieces
(or ``hopping''), respectively $\hat D$ and $\hat K$:
\beq
\hat H - \mu \hat N = \hat D + \hat K,
\eeq
where 
\beq
\hat D = \sum_i \left( 1 - \mu \right)(\hat n_{1,i} +  \hat n_{2,i} + \hat n_{3,i}) +  g \hat n_{1,i} \hat n_{2,i}  \hat n_{3,i},
\eeq
and $\hat K = \hat K_1 + \hat K_2 + \hat K_3$, where
\beq
\hat K_s = -\frac{1}{2}\sum_{i,j} \hat \psi^\dagger_{s,i} (\delta_{i,j+1} + \delta_{i,j-1} ) \hat \psi^{}_{s,j}.
\eeq

Armed with $\hat D$ and $\hat K$ as above, we may proceed in different ways. 
One way is to use a Suzuki-Trotter decomposition:
\beq
\label{Eq:TS}
e^{-\beta (\hat H -\mu \hat N)} \simeq e^{-\tau \hat K}e^{-\tau \hat D} \dots e^{-\tau \hat K}e^{-\tau \hat D} + \mathcal O(\beta^2/N_\tau),
\eeq
where $N_\tau$ factors are present on the right-hand side, such that $\beta = \tau N_\tau$.
The above decomposition can be expressed as a hopping-parameter expansion (see Appendix), 
\bea
\label{Eq:HPE}
&&\!\!\!\!\!\! e^{-\beta (\hat H -\mu \hat N)} \simeq
e^{-\beta \hat D} \sum_{m = 1}^{\infty} \sum_{k_{m} = 1}^{N_\tau} \sum_{k_{m-1} = 1}^{k_m-1}\dots \sum_{k_1 = 1}^{k_2-1}  \nonumber \\
&& e^{\tau k_m \hat D} ({-\tau \hat K})e^{-\tau k_m \hat D}\times \dots \times  e^{\tau k_1 \hat D} ({-\tau \hat K}) e^{-\tau k_1 \hat D}. \nonumber \\
\eea
The grand-canonical partition function corresponds to the trace of the above operator. Using a coordinate-space basis for all the operators involved, the trace can be written as a sum over configurations represented by lines, so-called worldlines, connecting the lattice points (see the description of the algorithm below). Each worldline represents how a given string of matrix products in Eq.~(\ref{Eq:HPE}) modifies a given basis wavefunction as it propagates in imaginary time. The worldlines are straight along the time direction unless an off-diagonal contribution $({-\tau \hat K})$ appears, in which case a worldline will deform sideways at that particular time slice. Because a trace is involved, all worldlines begin and end at the same spatial point when traversing the full extent $\beta = \tau N_\tau$ of the time direction. 

More specifically, the $\hat K_s$ factors are single-particle operators acting on each species independently, e.g.:
\bea
\bra{X} (-\tau \hat K_1) \ket{Y} =  \frac{\tau}{2} (\delta_{x_1,y_1 + 1} + \delta_{x_1,y_1 - 1}) \delta_{x_2,y_2} \delta_{x_3,y_3},
\eea
where we have used $\ket{Y} = \ket{y_1}  \ket{y_2}  \ket{y_3}$, and similarly for $\bra{X}$.
The kinetic energy operator will allow a worldline to move sideways between two neighboring 
spatial points on a given time slice.

The diagonal piece, on the other hand, is a combination of one- and three-body operators and must therefore
be treated differently. To that end, we write
\beq
e^{-\tau \hat D} = 
\prod_i  e^{-\tau k  \hat n_{1,i}} e^{-\tau k  \hat n_{2,i}} e^{-\tau k  \hat n_{3,i}} e^{- \tau g \hat n_{1,i}  \hat n_{2,i}  \hat n_{3,i}},
\eeq
where $k = 1 - \mu$ and moreover note that
\beq
e^{-\tau k  \hat n_{s,i}} = 1 + \left( e^{-\tau k} - 1 \right) \hat n_{s,i},
\eeq
which is diagonal in coordinate space and always equal to $e^{-\tau k} = e^{\tau (\mu - 1)}$; and 
\beq
e^{- \tau g \hat n_{1,i}  \hat n_{2,i}  \hat n_{3,i}} = 1 + \left(e^{-\tau g} -1\right) \hat n_{1,i}  \hat n_{2,i}  \hat n_{3,i}.
\eeq
which is also diagonal and always equal to 1 unless the initial and final states contain three non-identical particles that share the same location, in which 
case the interaction operates too and the matrix element equals $e^{-\tau g}$.
When the worldlines of all three flavors coincide at a given point in space and time, the interaction will operate and yield a nontrivial 
contribution to that particular term.

\subsection{Renormalization}
\begin{figure}[h]
\begin{center}
	\includegraphics[width=\linewidth]{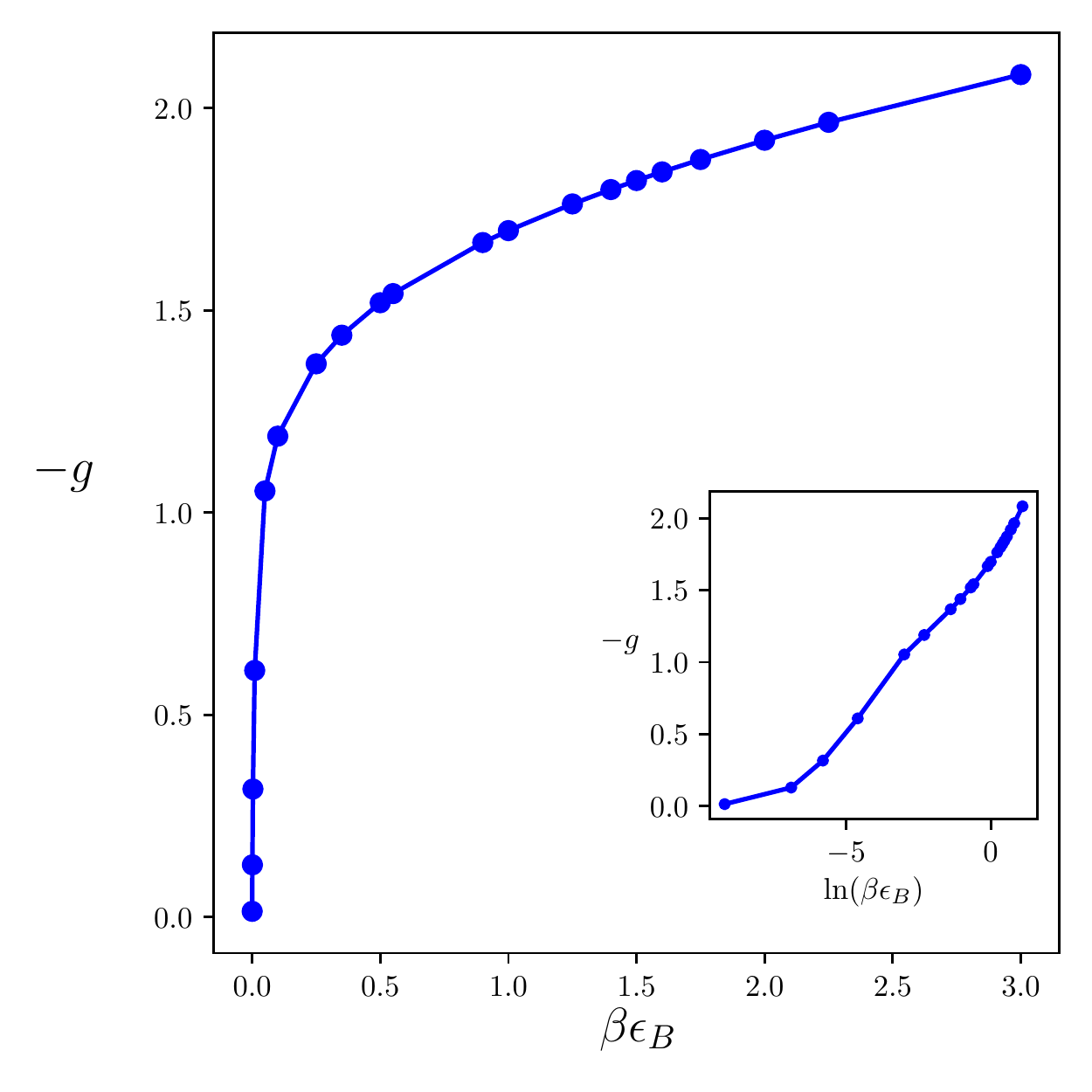}
	\caption{\label{Fig:coupling} The effective binding energy as a function of the coupling parameter $g$ for $N_{x}=80$.
	Inset: $-g$ as a function of $\ln (\beta \epsilon_B)$.}
\end{center}
\end{figure}

As the Hamiltonian~\eqref{Eq:H} features an ultraviolet divergence in 1D, the coupling $g$ must be renormalized~\cite{PhysRevLett.120.243002}.
In this work, however, as we employ open boundary conditions to avoid the infamous sign problem, the renormalization condition
of Ref.~\cite{PhysRevLett.120.243002} is inapplicable.
Instead, we numerically solve the three-body problem at each coupling strength along the lines of the iterative method presented
in Ref.~\cite{PhysRevA.99.013615}.
In particular, after expanding in a sine-wave basis, the value of the wave function for wavenumbers $n$, $m$, $l$ is
\beq\label{Eq:renorm}
\phi_{nml} = -\frac{8g/L^3}{\epsilon_{nml} + \epsilon_B} \sum_{i,j,k} \phi_{ijk} S^{ijk}_{nml},
\eeq
where the lattice length $L=N_x+1$, the kinetic energy is $\epsilon_{nml} = \epsilon_n + \epsilon_m + \epsilon_l - 3\epsilon_1$,
and the effective binding energy is $\epsilon_B = -(E-3\epsilon_1)$ (here, $\epsilon_k = 1 - \cos\left(k\pi/L\right)$,
and the ground-state kinetic energy $3\epsilon_1$ is subtracted to ensure $\epsilon_B \ge 0$).
The factor $S^{ijk}_{nml}$ is given by the overlap of the six sine-wave basis functions corresponding to its upper and lower indices
as wavenumbers:
\beq
S^{ijk}_{nml} = \sum_{x=1}^{N_x} \prod_q \sin\left(\frac{q \pi}{L} x\right),
\eeq
where $q \in \{i,j,k,n,m,l\}$.
For given values of $\epsilon_B$ and $L$, the value of $g$ which solves Eq.~\eqref{Eq:renorm} for all $n$, $m$, $l$ is
the renormalized coupling. The resulting relationship between $g$ and $\beta \epsilon_B$ is shown in Fig.~\ref{Fig:coupling}.
As expected, the dependence shows logarithmic behavior $-g \propto \ln (\beta \epsilon_B)$, for large enough $\beta \epsilon_B$.

As a final consideration of the lattice method, we now address dimensionless parameters and finite-size scaling.
To approach both the continuum and thermodynamic limits, we must satisfy the separation of scales
\beq
\ell \ll \lambda_T \ll L,
\eeq
where the thermal wavelength $\lambda_T = \sqrt{2\pi \beta}$ and, as mentioned previously, $\ell = 1$.
For fixed $\lambda_T/L$, a given (dimensionless) value of $\beta \epsilon_B$ corresponds to equivalent physical coupling strengths
across different lattice sizes; given a value $\lambda_T/L \ll 1$, we increase $\beta$ and $L$ until achieving
numerical convergence in the observables. 
The final dimensionless parameter, $\beta \mu$, controls the total number of particles in the system.

\section{Update algorithm and measurement of observables}

To carry out the Monte Carlo sampling of the partition function in the worldline representation [Eqs.~\eqref{Eq:Z} and~\eqref{Eq:HPE}],
we employ the ``worm'' algorithm as described in Ref.~\cite{PhysRevD.99.074511}, adapting it to the case of three fermion species with a three-body
interaction.

\begin{figure}[t]
\begin{center}
	\includegraphics[scale=0.31]{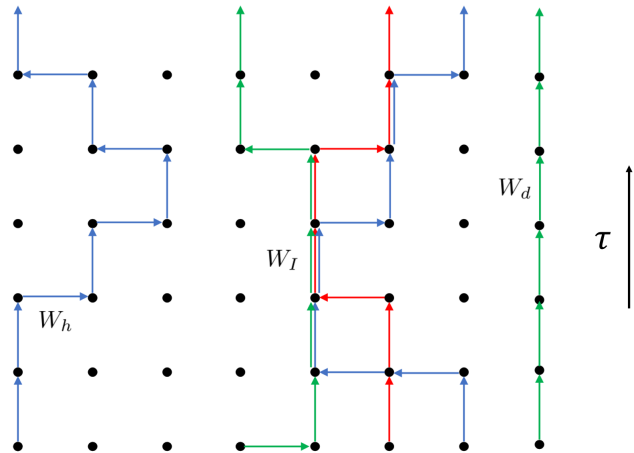}
	\caption{\label{Fig:Worldlines} Example configuration of worldlines contributing to Eq.~\eqref{Eq:HPE},
	where the three different colors correspond to the three fermionic species.
	Instances of the different weights are labelled where they occur; note that all worldlines are periodic
	in the temporal direction.
	}
\end{center}
\end{figure}

The worm algorithm evolves the system from one configuration $c$ of a set of worldlines (Fig.~\ref{Fig:Worldlines}) to another
by a series of contiguous local updates. The path traversed through the spacetime lattice by this
series of updates is known as the ``worm''; each individual local update occurs at the location of the ``head'' of the worm.
At the beginning of each global update, a spacetime lattice site is chosen at random; this site becomes the ``tail'' of the worm.
The head departs from the tail, effecting local updates as it meanders through the lattice.
The global update is complete when the head reunites with the tail, thereby closing a completed worldline or else annihilating
a previously existing one.

At each site visited by the head, the local update must obey detailed balance.
As transition probabilities that satisfy this requirement are tabulated in Ref.~\cite{PhysRevD.99.074511}, we do not reproduce them here;
we do, however, note that the relevant weights appear as
\begin{eqnarray}
W_d &=& e^{\tau\left(\mu-1\right)}, \\
W_h &=& \tau/2, \\
W_I &=& e^{-\tau g},
\end{eqnarray}
where $W_d$ corresponds to the diagonal term $\hat{D}$, $W_h$ to the hopping term $\hat{K}$, and
$W_I$ to the interacting contribution that occurs when three worldlines of different flavors share a time-like bond 
(wherever $W_I$ occurs, the three diagonal contributions $\left(W_d\right)^3$ are present as well). Instances of each of these weights
are illustrated in Fig.~\ref{Fig:Worldlines}.
For completeness, we provide a map of all possible\footnote{Sites on the spatial boundaries have restricted
sets of transitions, and the associated probabilities are modified; these are excluded from the Appendix due to their
negligible contribution in the large-volume limit.} local updates in the Appendix.

In terms of the weights listed above, the total weight of a spacetime lattice configuration $c$ is~\cite{PhysRevD.99.074511}
\beq
\Omega(c) = \left(W_d\right)^{N_{\tau} N} \left(W_h\right)^{n_h} \left(W_I\right)^{n_I},
\eeq
where $n_h$ is the total number of hops among all fermion species, and $n_I$ is the number of interacting time-like bonds
($n_I = 1$ in Fig.~\ref{Fig:Worldlines}).
Now, the partition sum may be written as
\beq\label{Eq:ConfigZ}
\mathcal{Z} = \sum_c \Omega(c),
\eeq
so that observables, generally given by 
\beq\label{Eq:O}
\braket{\hat{\mathcal{O}}} = \frac{1}{\mathcal{Z}} \sum_c \mathcal{O}(c)\, \Omega(c),
\eeq
may be computed by taking suitable log-derivatives of Eq.~\eqref{Eq:ConfigZ}.
As an example, the energy of a single configuration $c$ is calculated as
\beq
E(c) = N - \frac{n_h}{\beta} + g \frac{n_I}{N_{\tau}}.
\eeq
With configurations generated by the worm algorithm, Eq.~\eqref{Eq:O} is simply evaluated as the arithmetic mean of
the observable's value over the sampled configurations (after allowing for decorrelation between consecutive samples).

\section{Results}

We carried out our Monte Carlo calculations in lattices of spatial size $N_x = 80$ and temporal size $N_\tau = 3600$, with 
corresponding lattice spacings $\ell = 1$ and $\tau = 0.005$, such that $L = (N_x+1)\ell = 81$ (we used hard-wall boundary conditions
in the spatial directions) and $\beta = N_\tau \tau =18$.
We obtained data for 5 unique couplings, corresponding to $\beta \epsilon_B$ = $0.5$, $1.0$, 
$1.5$, $2.0$, $2.5$. At the time of sampling, particle density and energy were recorded, taking a total of $2\times 10^5$ 
de-correlated samples. Using these quantities and the appropriate Maxwell relations, we report other significant thermodynamic 
properties. Most values are reported as a ratio with respect to their noninteracting counterparts, as detailed below.

\subsection{Equations of state: Density, Energy, Pressure, and Contact}

\begin{figure}[t]
    \centering
    	\includegraphics[width=\columnwidth]{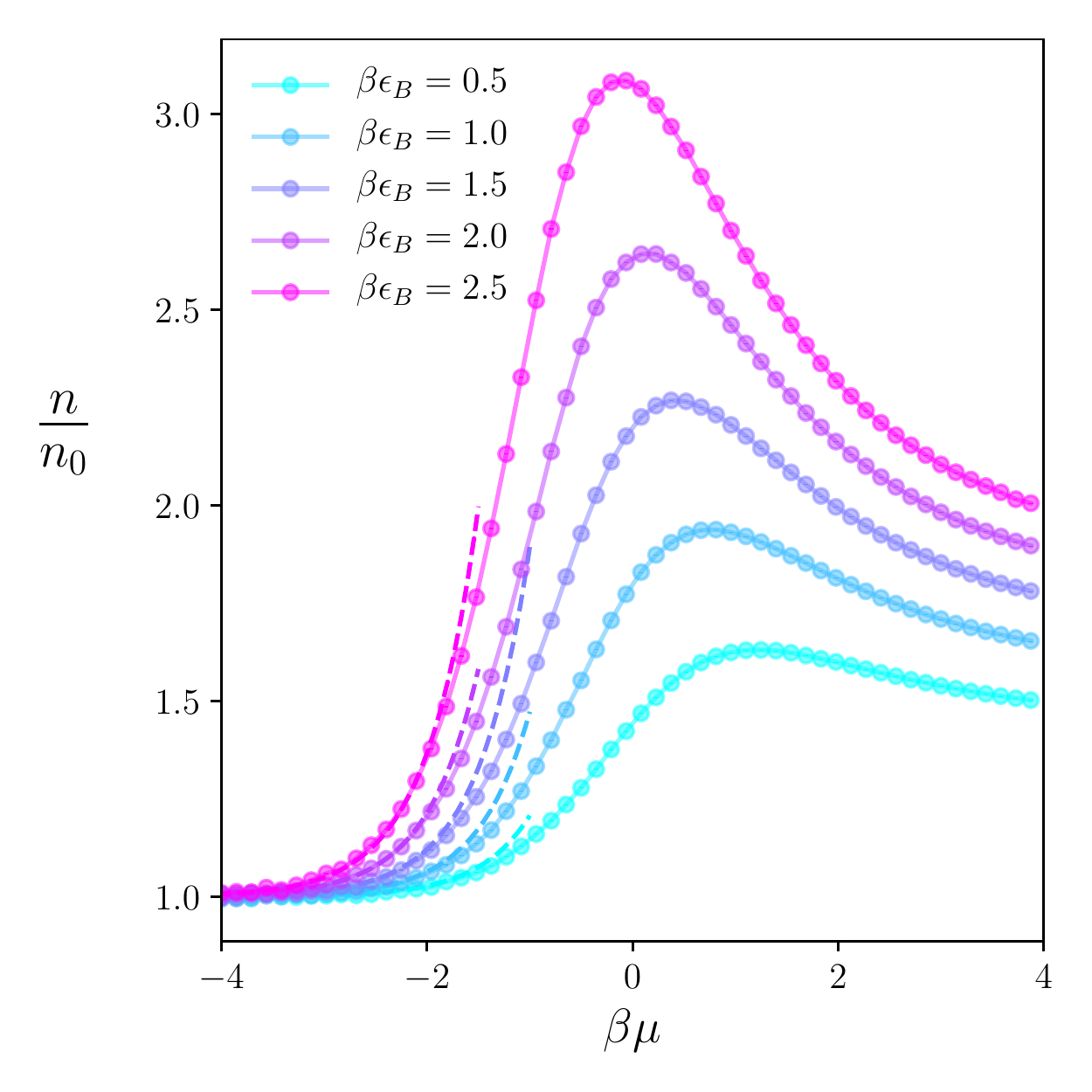}    
\caption{\label{Fig:Density} The density $n$ in units of the noninteracting density $n_0$ as a function of the dimensionless parameter $\beta \mu$ and the coupling strength $\beta \epsilon_{B}$. The solid lines interpolate the Monte Carlo results 
(shown with circles); the dashed lines show the third-order virial approximation.}
\end{figure}

The top panel of Fig.~\ref{Fig:Density} displays the interacting total density $n$ in units of its noninteracting counterpart $n_0$, as a function of $\beta \mu$, for several coupling strengths. The solid line interpolates the density for the values of $\beta \mu$ that were not explicitly calculated. For reference, we note that the total noninteracting density is derived from the Fermi-Dirac distribution,
\beq
\label{eq:nonint}
n_0 = \frac{3}{L}\sum_{k = 1}^{N_x} \frac{e^{-\beta(\epsilon_k - \mu)}}{1 + e^{-\beta(\epsilon_k - \mu)}},
\eeq
where $\epsilon_k = 1-\cos\left(k\pi/L\right)$ is simply the Hubbard dispersion relation resulting from nearest-neighbor hopping
(i.e. a double-difference formula for the second derivative).

In the same figure, we compare our results to the third-order virial expansion
\beq
\label{Eq:VirialDensity}
n \lambda_{T} / 3 = z + 2b_2^0 z^{2} + 3(b_3^0 + \Delta b_3)z^3,
\eeq
where $\lambda_T = \sqrt{2\pi \beta}$, $z = e^{\beta \mu}$ is the fugacity and $b_{i}^{0}$ is the $i$-th noninteracting virial coefficient. As shown in Ref.~\cite{PhysRevLett.120.243002}, $\Delta b_3$ is related to the second-order virial coefficient 
$\Delta b_2^{\mathrm{2D}}$ of 2D, two-species fermions by
\begin{equation}
\Delta b_3 = \frac{1}{\sqrt{3}} \Delta b_2^{\mathrm{2D}},
\end{equation}
where
\begin{equation}
\Delta b_2^{\mathrm{2D}}(\beta \epsilon_B) = \nu(\beta \epsilon_B) = \int_0^{\infty} dt\ \frac{\left(\beta \epsilon_B\right)^t}{\Gamma(t+1)}.
\end{equation}

The density equation of state $n/n_0$ displays the same characteristic behavior observed both in 1D and 2D Fermi systems with two-body interactions: a rapid healing to the virial expansion at negative $\beta \mu$; a maximum where quantum fluctuations dominate, typically around $\beta \mu = 0$; and a relatively mild decay at large and positive $\beta \mu$ where the increased density tends to partially quench interaction effects.

\begin{figure}[t]
    \centering
	\includegraphics[width=\columnwidth]{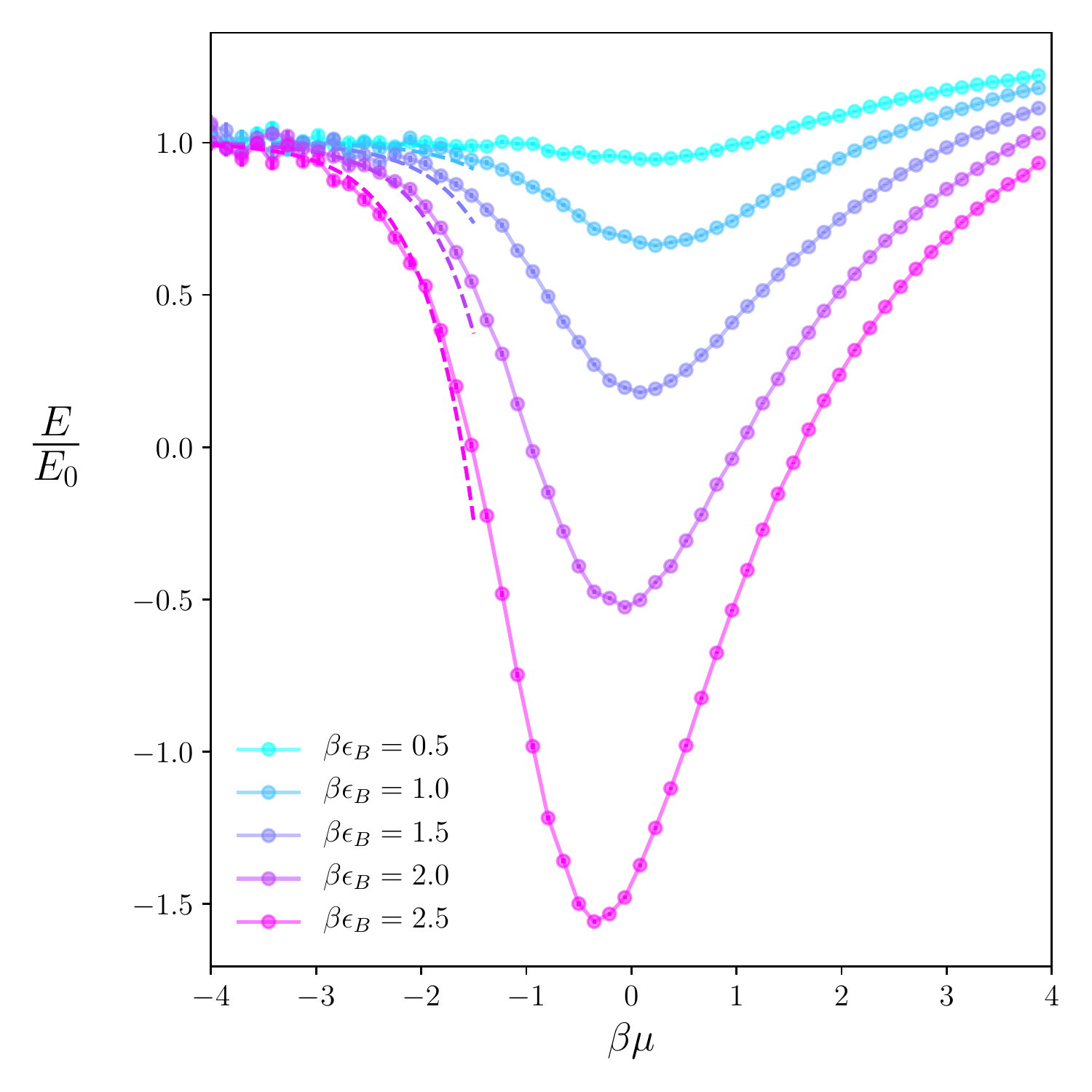}    
\caption{\label{Fig:Energy} The total energy $E$ in units of the noninteracting energy $E_0$ as a function of $\beta \mu$, for 
several values of the coupling $\beta \epsilon_{B}$ (same values as in Fig.~\ref{Fig:Density}). The solid lines interpolate the Monte Carlo results 
(shown with circles); the dashed lines show the third-order virial approximation.
}
\end{figure}

In addition to the density, we calculated the energy equation of state across multiple chemical potentials and couplings, as shown in the bottom panel of Fig.~\ref{Fig:Energy}. There, the energy $E = \langle \hat H \rangle$ is shown in units of its noninteracting value given by
\beq
E_0 =  3\sum_{k = 1}^{N_x} \frac{\epsilon_k \; e^{-\beta(\epsilon_k - \mu)}}{1 + e^{-\beta(\epsilon_k - \mu)}}.
\eeq
The energy equation of state $E/E_0$ displays features with obvious counterparts in the density described above. 
Indeed, $E/E_0$ heals to the virial expansion at large negative $\beta \mu$; the interplay of quantum and interaction fluctuations results in a broad minimum around $\beta \mu = 0$ for all couplings; and finally, at large positive $\beta \mu$, the interaction effects are progressively softened. 
%

The analysis of Refs.~\cite{PhysRevLett.120.243002,PhysRevA.99.013615} suggested that this system undergoes a ``Fermi-Fermi crossover'' in moving from weak to strong coupling,
where, starting from three species of ideal Fermi gases, increasing attraction causes the formation of a repulsive Fermi gas of composite trimers.
In the limit of infinite coupling strength, we anticipate that as the trimer radius shrinks to zero,
the repulsion between trimers will give way to ideal behavior, so that the trimers behave as a hard-core Bose gas in 1D~\cite{girardeau}.
With our many-body results, we may shed additional light on this question.

\begin{figure}[t]
\centering
\includegraphics[width=\linewidth]{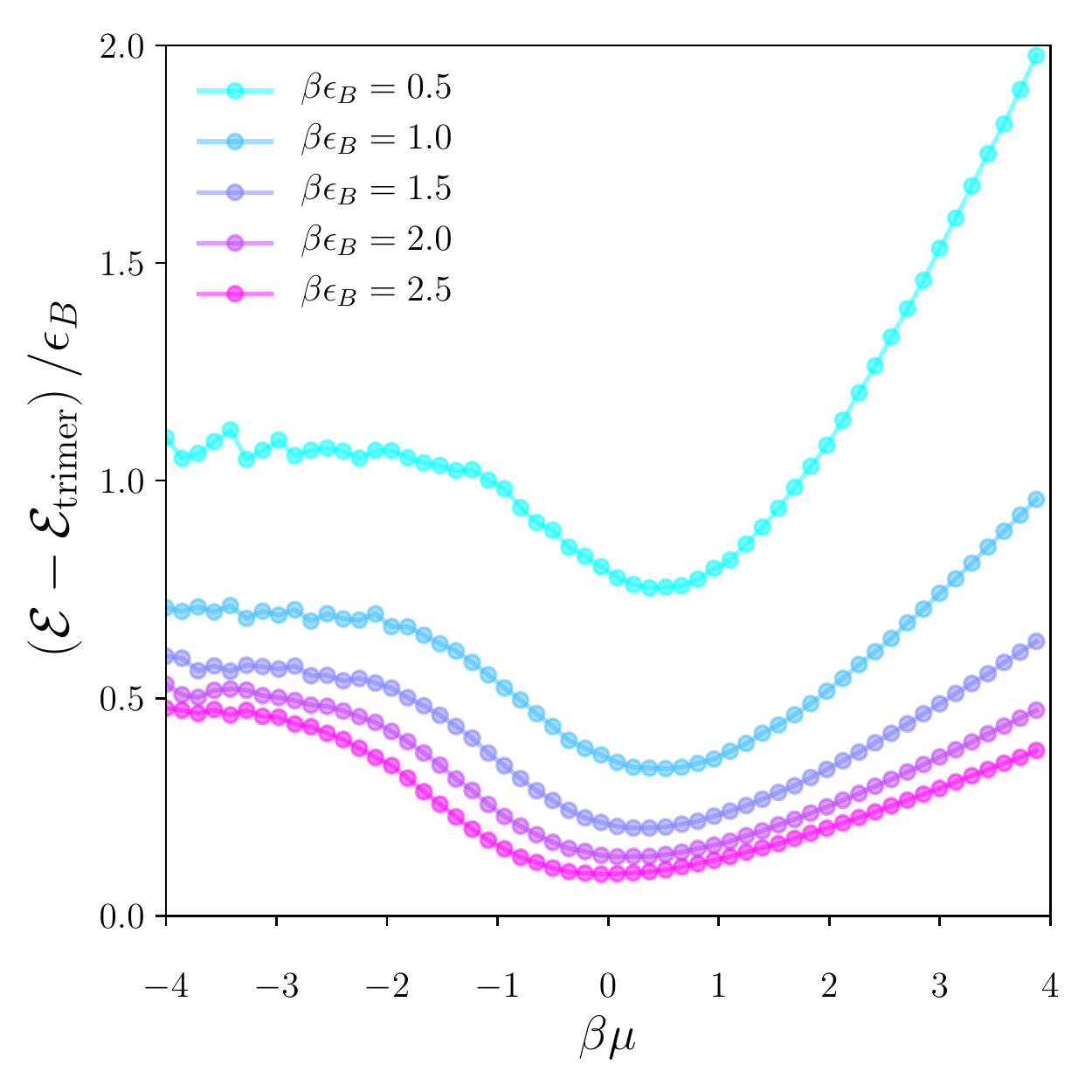}    
\caption{\label{Fig:trimer_gas} Difference in energy per particle $\mathcal E = E/N$ (in units of $\epsilon_B$) between the interacting system and the
energy per particle $\mathcal E_\text{trimer}$ of the corresponding ideal gas of composite trimers (with the same total particle number; see text).}
\end{figure}

To compare our results with an ideal Fermi gas of trimers, we calculate the energy of the trimer gas as follows.
For a given value of $\beta$ and $\mu$, the trimer chemical potential $\mu_{\mathrm{trimer}}$ is tuned so that the total number of particles
is equal in both systems: $3N_{\mathrm{trimer}} = N$.
The trimer energy is then computed as $E_{\mathrm{trimer}} = \expval{E}_{\mathrm{trimer}} - \epsilon_B N/3$,
where the expectation value is that of an ideal Fermi gas with mass $3m$.
In Fig.~\ref{Fig:trimer_gas}, the difference in energy per particle $\mathcal E = E/N$ between our results and the ideal trimer gas is displayed,
showing that increasing the coupling strength does indeed cause the system to approach the trimer gas.
The positivity of each of the curves supports the prediction of Ref.~\cite{PhysRevA.99.013615} that, 
for finite coupling, the effective interaction between trimers is repulsive.

\begin{figure}[t]
\centering
\includegraphics[width=\linewidth]{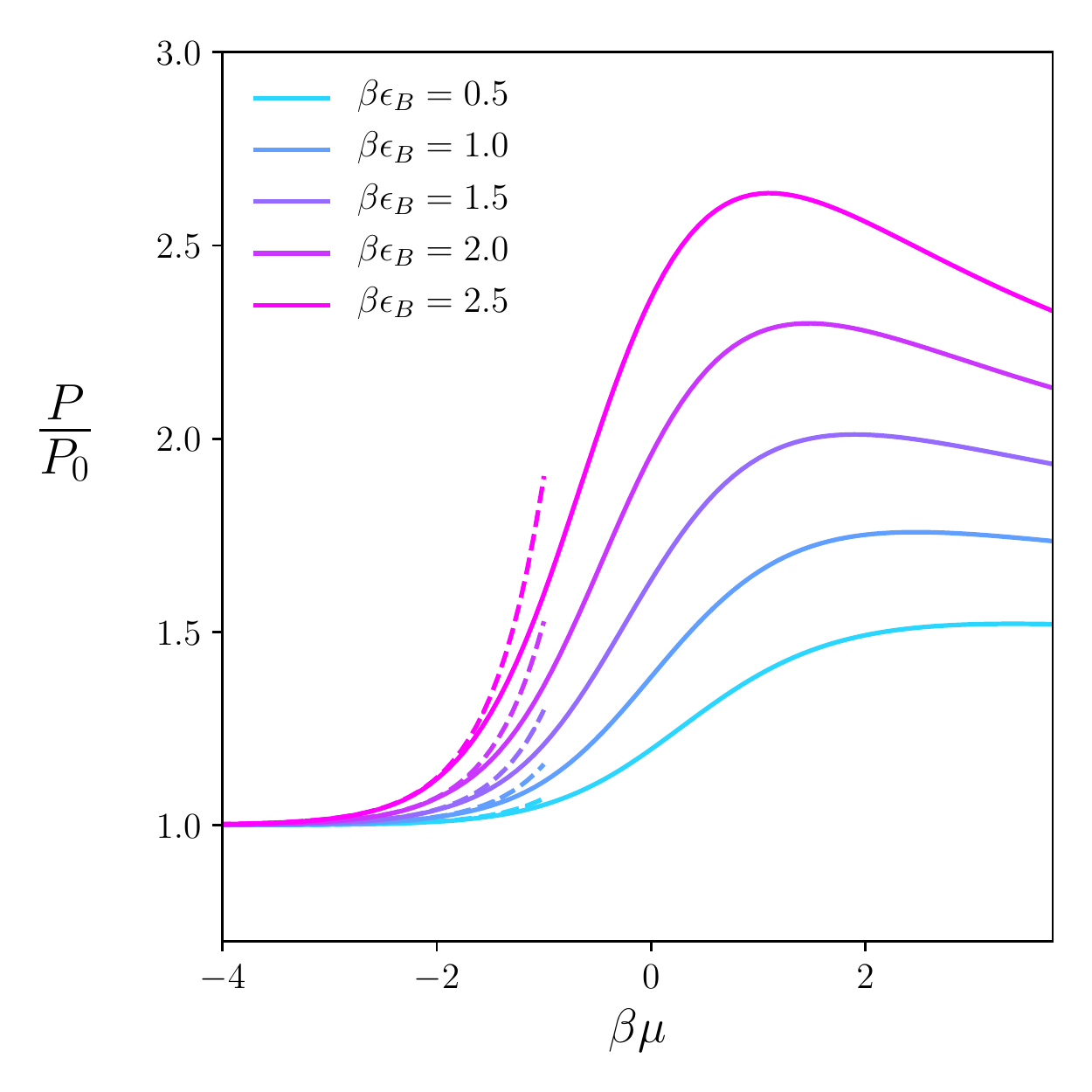}    
\caption{ \label{Fig:pressure} The pressure $P$ in units of the noninteracting pressure $P_0$ as a function of $\beta \mu$,
obtained by numerical integration of the density equation of state. The dashed line is the result from the third-order virial approximation.}
\end{figure}

From the density equation of state, the pressure can be obtained by integration, namely
\beq
\label{eq:pressure}
P \lambda_T^{3} = 2\pi \int_{-\infty}^{\beta \mu} \dd(\beta \mu)' n \lambda_T .
\eeq
For reference, we note that the noninteracting pressure is given by
\beq
P_0 =  \frac{3}{\beta L} \sum_{k = 1}^{N_x} \ln\left(1 + e^{-\beta(\epsilon_k - \mu)}\right).
\eeq
Our results using the above expressions and numerical integration are shown in Fig.~\ref{Fig:pressure}. To carry out the
integration, the virial expansion was used as a boundary condition at large negative $\beta \mu$; specifically,
the virial approximation of Eq.~(\ref{Eq:VirialDensity}) was used for $\beta \mu < 4$ in the integrand of Eq.~\eqref{eq:pressure}.

Tan's contact density, which governs the behavior of correlation functions at short distances~\cite{TAN20082952, TAN20082971, TAN20082987}, can be calculated in our approach via
\beq
\mathcal C = \frac{1}{L}\frac{\partial \ln \mathcal Z}{\partial (\beta \epsilon_B)} = 
\frac{1}{L}\frac{\partial \ln \mathcal Z}{\partial g} \frac{\partial g}{\partial (\beta \epsilon_B)} = 
-\frac{\beta \langle \hat V \rangle}{L} \frac{\partial \ln g}{\partial (\beta \epsilon_B)},
\eeq
where the $\langle \hat V \rangle$ factor encodes the many-body aspects of the problem, whereas 
${\partial \ln g}/{\partial (\beta \epsilon_B)}$ is fully determined by the renormalization of $g$, i.e. purely three-body physics.
\begin{figure}[h]
\centering
\includegraphics[width=\linewidth]{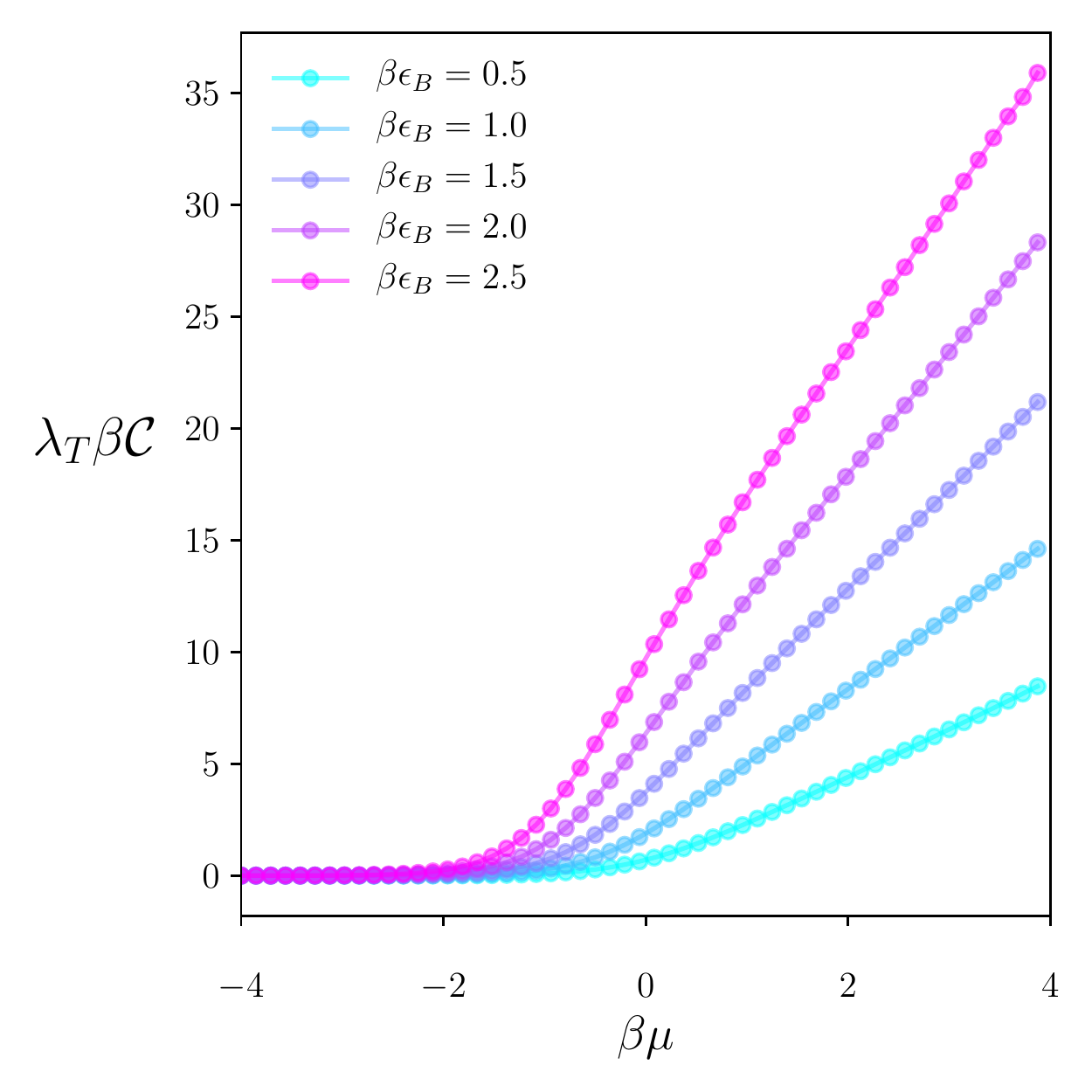}    
\caption{\label{Fig:contact} Contact density $\mathcal C$ obtained from the expectation value of the potential energy, as a function
of $\beta \mu$ and for several values of the coupling strength.}
\end{figure}
In Fig.~\ref{Fig:contact} we show our results for $\mathcal C$ as a function of $\beta \mu$ and for several values of the 
coupling strength. The contact can also be obtained from the pressure-energy relation, which in 1D takes the form 
$P - 2E/L = \mathcal C$. However, the latter relies on non-relativistic invariance, which is broken by lattice
artifacts, and therefore yields somewhat different results at strong coupling or large densities (see Appendix).

\begin{figure}[h]
\centering
\includegraphics[width=\linewidth]{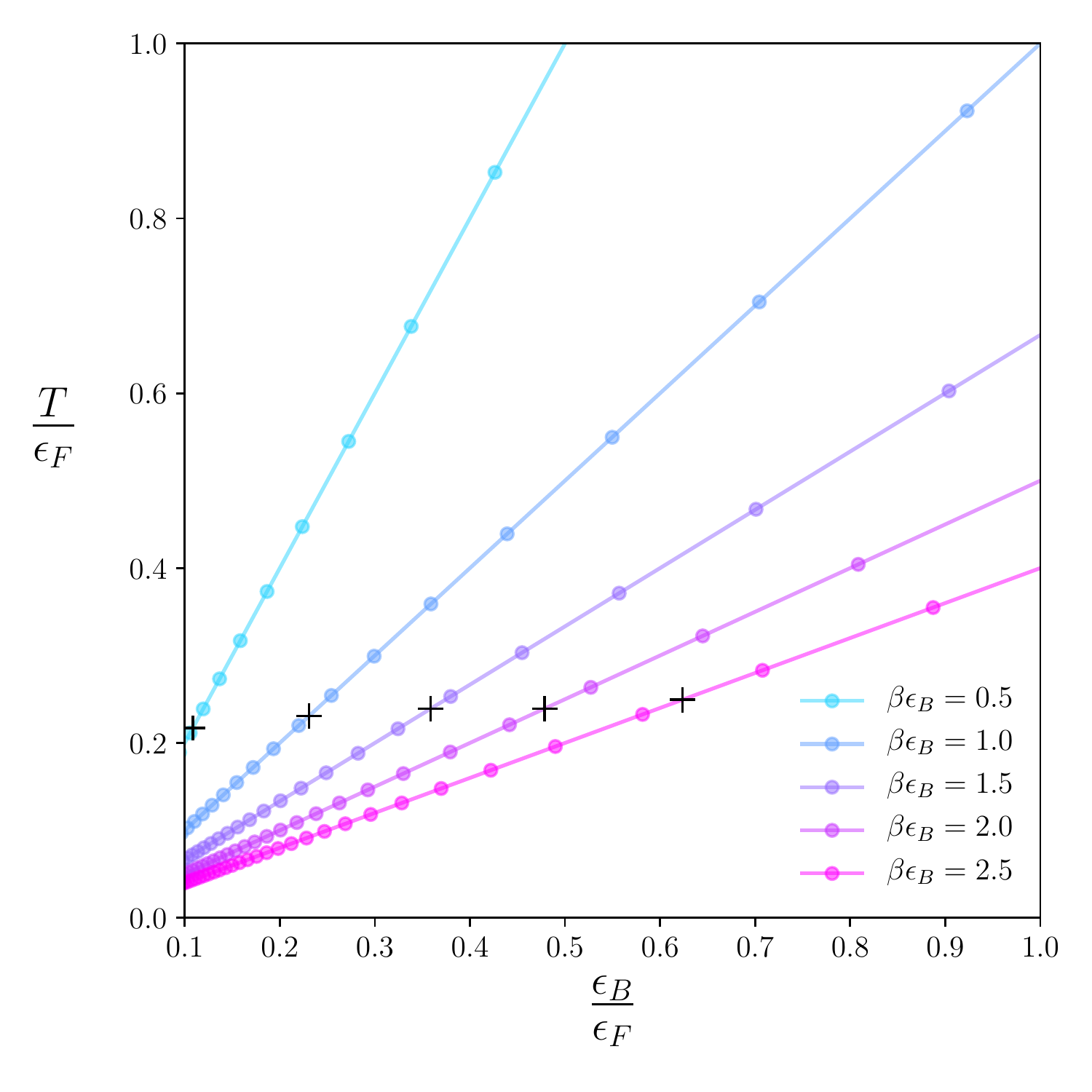}    
\caption{\label{Fig:tempscale}The temperature scale $T/\epsilon^{}_F$ as a function of the binding energy $\epsilon^{}_B /\epsilon^{}_F$,
where $\epsilon^{}_F = n^2$, for each of the couplings studied. The solid lines interpolate our Monte Carlo results shown in circles.
The crosses mark the location of the maxima in $n/n_0$ at the corresponding
coupling strength (see Fig.~\ref{Fig:Density}).}		
\end{figure}

Finally, we note that, using the density equation of state shown above, one may connect our results as a function of 
$\beta \mu$ with the more conventional temperature scale $T/\epsilon_F$, where $\epsilon_F = (n/3)^2$,
as well as with the more conventional coupling scale $\epsilon_B /\epsilon_F$.
To that end, we display in Fig.~\ref{Fig:tempscale} the quantity $T/\epsilon_F$ as a function of $\epsilon_B /\epsilon_F$,
for each value of $\beta \epsilon_B$ we studied. From this plot we conclude, for instance, that the largest values of 
$\beta \epsilon_B$, for a given $\epsilon_B /\epsilon_F$, produce the lowest temperatures $T/\epsilon_F$.

\subsection{Static response functions: compressibility and magnetic susceptibilities}

As the three fermion species' chemical potentials may each be tuned independently, there are three independent magnetization 
(or polarization) parameters. By using scaled Jacobi coordinates, these parameters may be expressed in terms of previously 
defined quantities as follows. The first is 
\beq
\hat M_1 = \hat N = \hat N_1 + \hat N_2 + \hat N_3, 
\eeq
and corresponds to the total particle number and thus reflects compressibility effects (see below), rather than polarization.
The other two capture differences in particle content among the three flavors and quantify the degree to which the system is polarized:
\bea
\hat M_2 &=& \frac{3}{2}\left(\hat N_1 - \hat N_2\right), \\ 
\hat M_3 &=& \frac{1}{2}\left(\hat N_1 + \hat N_2\right) - \hat N_3, 
\eea
where overall factors are unimportant. The only requirement is that the applied fields $h_i$, defined in terms of the $\mu_i$, 
leave invariant the inner product
\beq
\sum_i h_i \hat M_i = \sum_i \mu_i \hat N_i.
\eeq

Thus, we have isothermal susceptibilities associated with each $\hat M^{}_i$ and given by
\beq
\chi^{}_i = \beta \left(\expval{\hat M_i^2} - \expval{\hat M^{}_i}^2 \right),
\eeq
where the polarizations $\hat M_i$ may be expressed in terms of the particle numbers $\hat N_i$ as shown above.

For balanced systems where $\mu_1 = \mu_2 = \mu_3 = \mu$, so that $h_2 = h_3 = 0$, $h_1 = \mu$,
$\chi_2$ and $\chi_3$ take on the same form when expressed in terms of their noninteracting counterparts:
\beq
\frac{\chi^{}_2}{\chi^0_2} = \frac{\chi^{}_3}{\chi^0_3} = \frac{\expval{\hat N_i^2} - \expval{\hat N^{}_i \hat N^{}_j}}{\expval{\hat N_k^2}^{}_0 - \expval{\hat N^{}_k \hat N^{}_l}^{}_0},
\eeq
where $i\ne j$, $k\ne l$, and $\expval{\cdot}_0$ denotes the noninteracting expectation value.

Note that, to determine the isothermal compressibility $\kappa$, we study the fluctuations in particle number:
\begin{equation}
	\expval{\hat N^2} - \expval{\hat N}^2 = \frac{1}{\beta}\left(\pdv{N}{\mu}\right)_{T,V},
\end{equation}
such that
\begin{equation}
\kappa = \frac{L \beta\left(\expval{\hat M_1^2} - \expval{\hat M^{}_1}^2\right)}{\expval{\hat M^{}_1}^2} = \frac{L \chi^{}_1}{\expval{\hat M^{}_1}^2}.
\end{equation}

\begin{figure}[h]
\centering
\includegraphics[width=\linewidth]{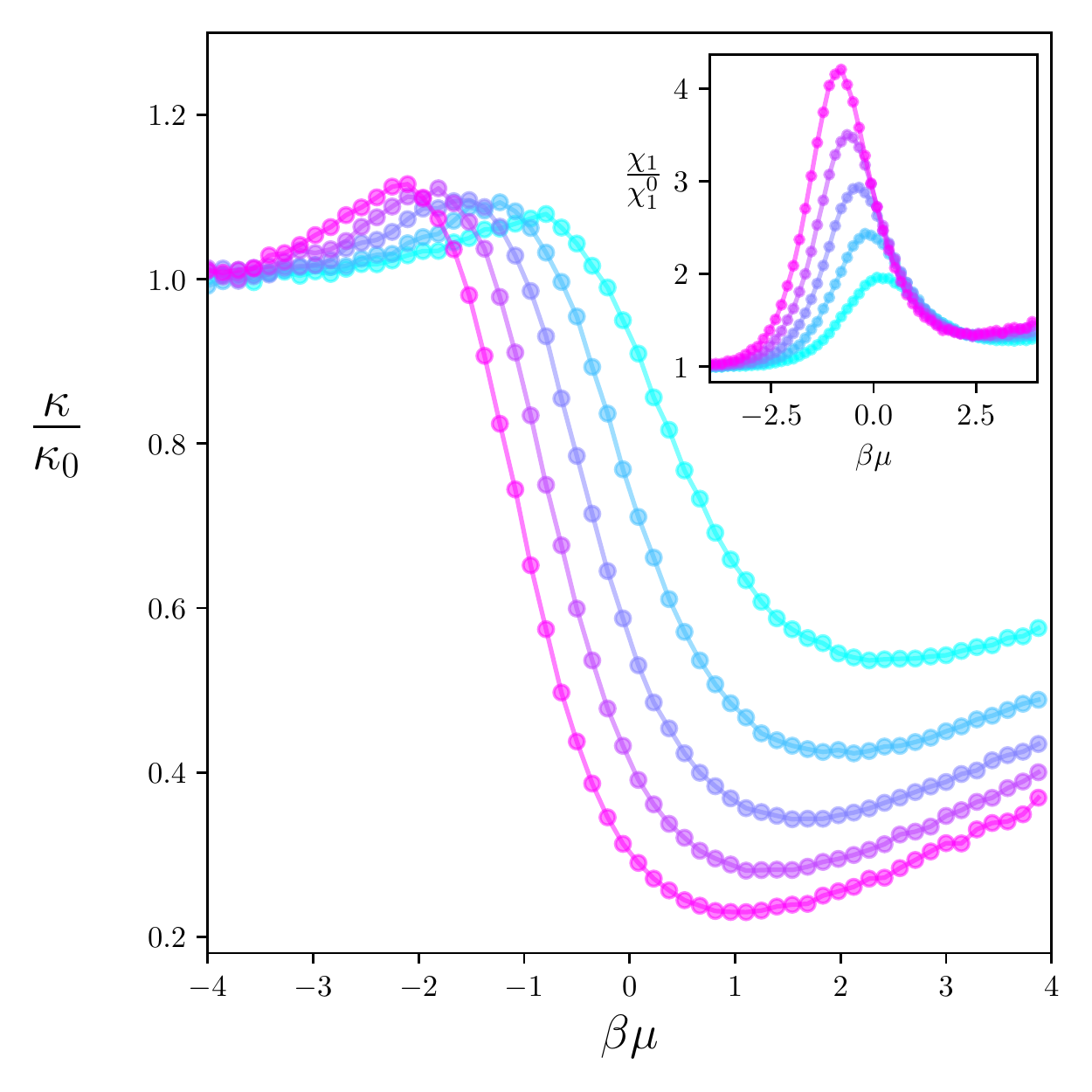}    
\caption{\label{Fig:compress} Compressibility $\kappa$ in units of its noninteracting value $\kappa^{}_0$, as a function of $\beta \mu$, for several values of the coupling $\beta \epsilon^{}_B$. The solid lines interpolate our Monte Carlo results shown in circles.
Inset: $\chi_1/\chi_1^{0}$ as a function of $\beta \mu$.}
\end{figure}

In Fig.~\ref{Fig:compress} we show $\kappa/\kappa^{}_0$, where $\kappa^{}_0$ is the compressibility of the noninteracting system, given by
\beq
\kappa_0 = \frac{3\beta}{4 L n_0^2} \sum_{k=1}^{N_x} \cosh^{-2}\left[\frac{\beta}{2}\left(\epsilon_k-\mu\right)\right],
\eeq
with $n_0$ from Eq.~\eqref{eq:nonint}.
Although not apparent in that plot, the interaction effects on the density fluctuations do attain a broad maximum as $\beta \mu = 0$ is approached, as shown explicitly in the inset of the same figure.

\begin{figure}[h]
    \centering
    \includegraphics[width=\linewidth]{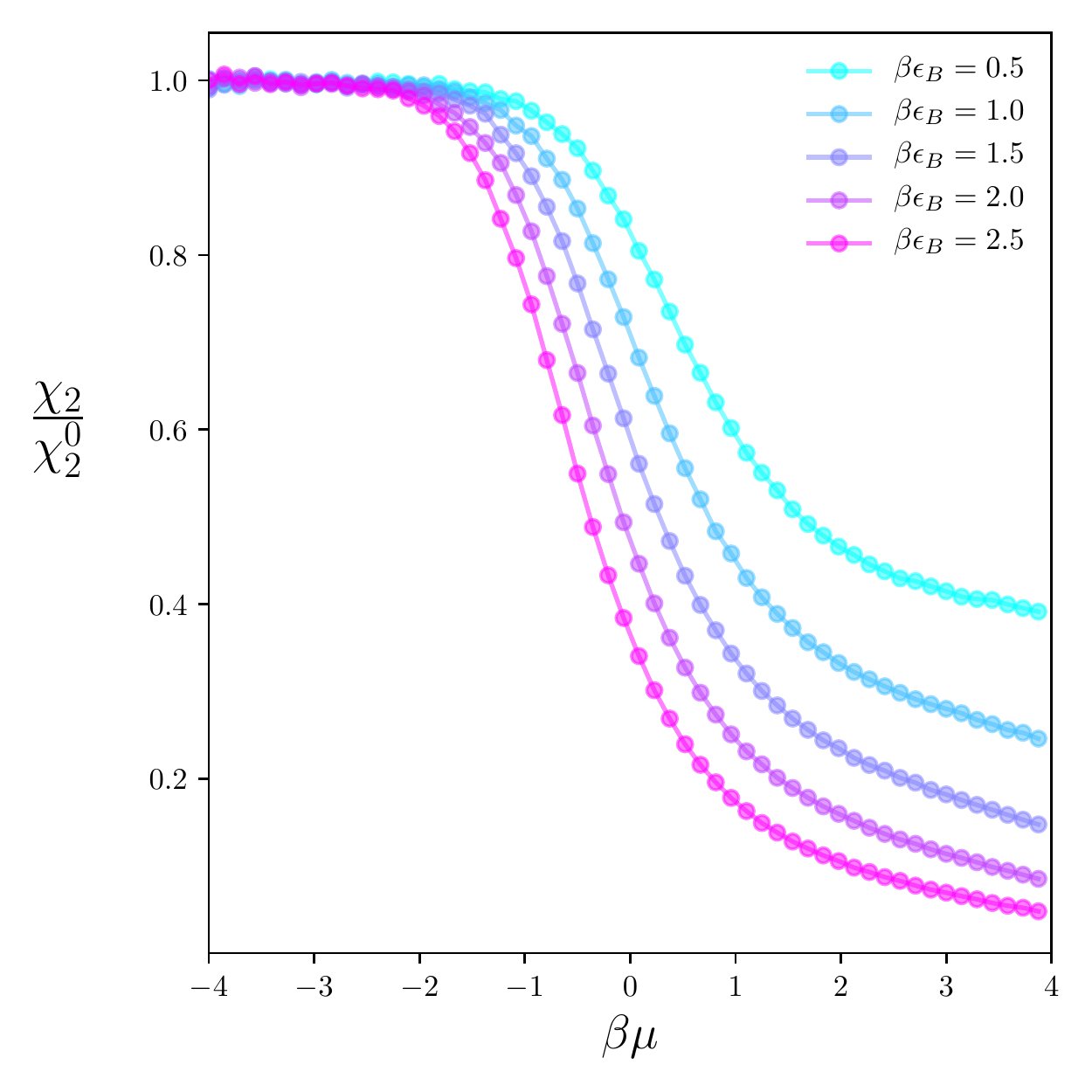}    
\caption{The interacting susceptibility $\chi_2$, in units of its noninteracting counterpart $\chi_2^{0}$, as a function of $\beta \mu$.
The solid lines interpolate our Monte Carlo results shown in circles.}
\label{Fig:magnetic}
\end{figure}

\section{Summary and Conclusions}

In this work we have implemented a non-perturbative stochastic method to investigate the properties of
nonrelativistic fermions in 1D with attractive three-body interactions. As previous works have noted, this system 
displays a scale anomaly: it is classically scale invariant in the sense that the interactions do not introduce a new scale
in the Hamiltonian, but quantum fluctuations generate a three-body bound state and thus scale-invariance is 
spontaneously broken.
There is, furthermore, an exact correspondence between the three-body sector of this system and the two-body
sector of the analogue system in 2D with two-body interactions. However, that correspondence is limited to those
specific sectors and does not extend to higher particle numbers or thermodynamics (certainly not easily). Therefore,
here we set out to provide a first characterization of the thermodynamics.

Using the worldline representation of the partition function, and using hard-wall boundary conditions, the sign problem
can be completely avoided in this 1D system, even though the number of species is odd. With that approach, supplemented
by the worm algorithm, we have calculated several equations of state: density, energy, and pressure, as well
as Tan's contact. Additionally, we have characterized the static thermodynamic response by calculating the
compressibility and the magnetic susceptibility. Our findings support the notion of a continuous Fermi-Fermi crossover,
originally pointed out and explored in Refs.~\cite{PhysRevLett.120.243002, PhysRevA.99.013615}, whereby
a trimer gas with repulsive interactions is approached at low temperatures, high-densities, or strong couplings.

As with similar systems with short-range interactions in 1D~\cite{PhysRevA.91.033618} and 2D~\cite{PhysRevLett.115.115301}, 
the interaction effects on the static response functions are maximized in the vicinity of $\beta \mu = 0$. Indeed, at large negative
$\beta \mu$, the system is very dilute and governed by few-particle physics (i.e. the virial expansion), where kinetic energy
contributions typically dominate; on the other hand, at large positive $\beta \mu$, high densities will tend to reduce the available
phase space and thus quench interaction effects.

\section{Acknowledgements}

We thank H. Singh and S. Chandrasekharan for helpful discussions and comments on the manuscript. 
This work was supported by the U.S. National Science Foundation under 
Grant No. PHY1452635.

\appendix
\section{Hopping-parameter expansion}

In this appendix we derive the hopping-parameter expansion of Eq.~(\ref{Eq:HPE}) by two methods:
starting from a lattice formulation, and starting from a continuum formulation (and discretizing at the end).

For the first approach, starting from the Suzuki-Trotter decomposition
\beq
e^{-\beta (\hat H -\mu \hat N)} \simeq e^{-\tau \hat K} e^{-\tau \hat D} \dots e^{-\tau \hat K} e^{-\tau \hat D}
+ \mathcal O(\beta^2 /N_\tau),
\eeq
with $\beta = \tau N_\tau$, we proceed by expanding the $e^{-\tau \hat K}$ factors in a power series:
\bea
e^{-\beta (\hat H -\mu \hat N)} &\simeq& \sum_{k_j = 0}^{\infty} \frac{({-\tau \hat K})^{k_{N_\tau}}}{k_{N_\tau}!}e^{-\tau \hat D} \dots 
\frac{({-\tau \hat K})^{k_1}}{k_1!}e^{-\tau \hat D}. \nonumber \\
\eea
We then organize the above multiple sum according to the value of
$n = \sum_j k_j$. The $n=0$ order simply yields $e^{-\beta \hat D}$. For $n=1$ we obtain
\beq
e^{-\beta \hat D} \sum_{k_1 = 1}^{N_\tau} e^{\tau k_1 \hat D} ({-\tau \hat K}) e^{-\tau k_1 \hat D}.
\eeq
For $n=2$ we have two kinds of contributions: those which apply $\hat K$ twice at the same time slice [$N_\tau$ terms, each with an 
extra factor of $1/2!$]; and those which apply $\hat K$ once at two different time slices [$N_\tau (N_\tau - 1)/2$ terms].
Both of these kinds of terms are $\mathcal O (\tau^2)$, but the first kind scale only as $N_\tau$ and will therefore drop out 
at large $N_\tau$ (i.e. in the small $\tau$ limit, holding $\beta$ fixed). [Another way of saying this is that the diagonal terms will form a set of measure zero 
in the continuum limit.]  Thus, the result for $n=2$ is 
\beq
e^{-\beta \hat D} \sum_{k_2 = 1}^{N_\tau} \sum_{k_1 = 1}^{k_2-1} e^{\tau k_2 \hat D} ({-\tau \hat K}) e^{-\tau (k_2-k_1) \hat D} ({-\tau \hat K}) e^{-\tau k_1 \hat D}.
\eeq
Proceeding in this fashion order by order in $n$, and summing over $n$ from $0$ to $\infty$, we obtain Eq.~(\ref{Eq:HPE}) in the main text.

For the second approach, we seek an expression for the evolution operator
\beq
\hat U(\beta) = e^{- \beta (\hat D + \hat K)} = e^{-\beta \hat D} \hat {\mathcal O}(\beta),
\eeq
where we have defined the imaginary-time-dependent operator
\beq
\hat {\mathcal O}(\tau) \equiv e^{\tau \hat D}e^{- \tau (\hat D + \hat K)},
\eeq
whose time derivative is
\beq
\label{eq:diffeq}
-\frac{\partial \hat {\mathcal O}(\tau)}{\partial \tau} =  \hat K_I(\tau) \hat {\mathcal O}(\tau),
\eeq
where
\beq
\label{eq:KI}
\hat K_I(\tau) \equiv e^{\tau \hat D} \hat K  e^{-\tau \hat D}.
\eeq

Integrating Eq.~(\ref{eq:diffeq}) with $\hat {\mathcal O}(0)  = \mathds{1}$ as the 
initial condition, we arrive at the integral equation
\beq
\label{eqn:fundThm}
\hat {\mathcal O} (\beta) =  \mathds{1} + \int_0^{\beta} d\tau\ \frac{\partial \hat {\mathcal O}(\tau)}{\partial \tau}
= \mathds{1} - \int_0^{\beta} d\tau \hat K_I(\tau) \hat {\mathcal O}(\tau) .
\eeq
Iteratively substituting Eq.~\eqref{eqn:fundThm} into itself generates a Dyson series,
\begin{multline}\label{eqn:dyson}
\hat {\mathcal O}(\beta) = \bigg[ \mathds{1} - \int_0^{\beta} d\tau_1\ \hat{K}_I(\tau_1) \\
+ \int_0^{\beta} d\tau_1 \int_0^{\tau_1} d\tau_2\ \hat{K}_I(\tau_1) \hat{K}_I(\tau_2) - \ldots \bigg].
\end{multline}
To evaluate the partition function $\mathcal{Z} = \mathrm{Tr} \left[e^{-\beta \left(\hat{D}+\hat{K}\right)}\right]$, 
then, it suffices to evaluate the trace of $\hat U(\beta)$ using the above expression for $\hat {\mathcal O}(\beta)$.
Doing that and expanding the $\hat{K}_I$ terms using Eq.~(\ref{eq:KI}), we obtain
\begin{multline}
\mathcal{Z} = \sum_{k=0}^{\infty} \int_0^{\beta} d\tau_k \int_0^{\tau_k} d\tau_{k-1} \ldots \\
\int_0^{\tau_2} d\tau_1\
\mathrm{Tr}\left[ e^{-(\beta-\tau_k)\hat{D}} \hat{K} e^{-(\tau_k-\tau_{k-1})\hat{D}} \hat{K} \ldots \hat{K} e^{-\tau_1 \hat{D}} \right],
\end{multline}
which yields Eq.~\eqref{Eq:HPE} upon discretization.

\section{Map of local updates}

\begin{figure}[b]
\begin{center}
	\includegraphics[width=\columnwidth]{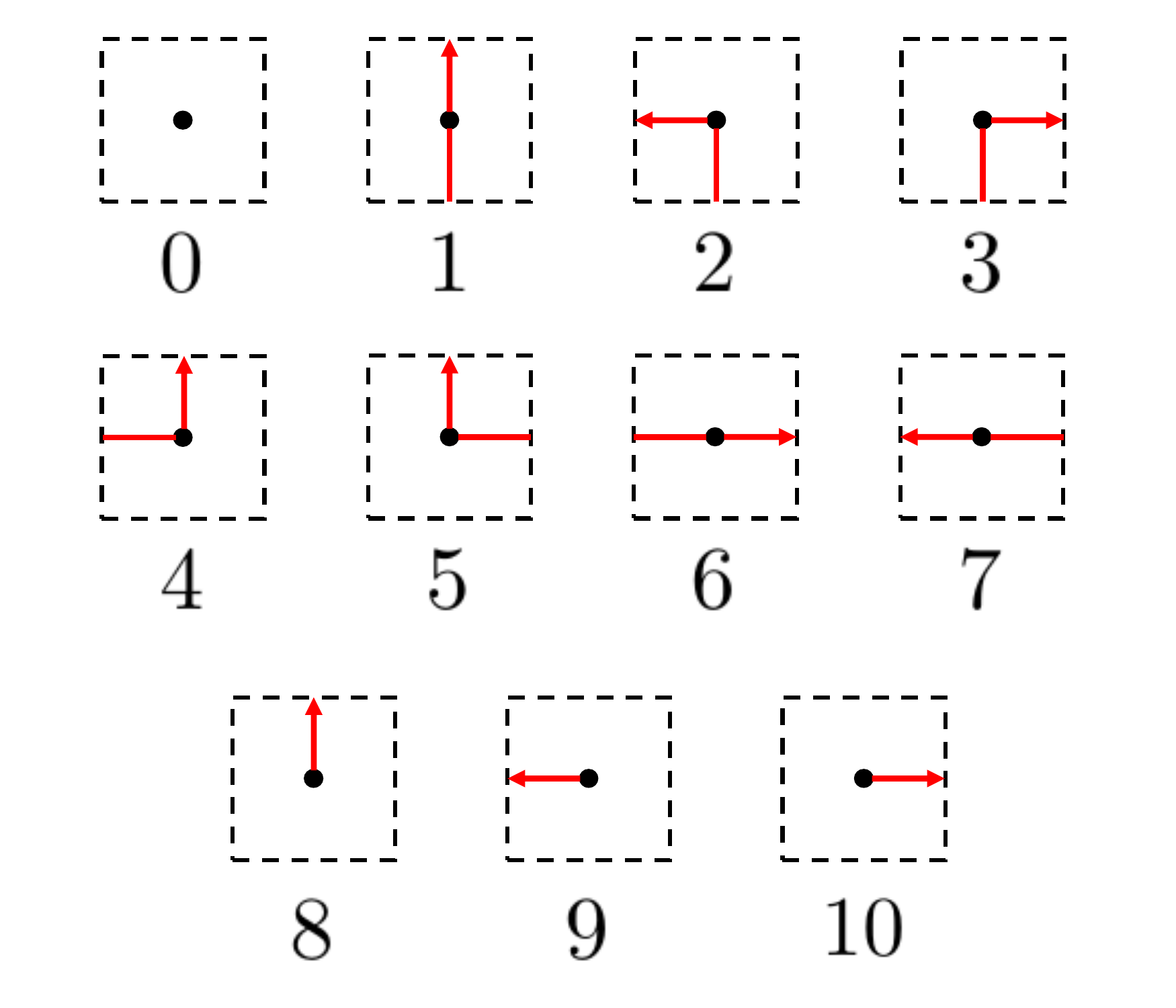}
	\caption{\label{Fig:Tiles} Local configurations of worldlines, with labels corresponding to those used in Table~\ref{tab:Tiles}.
	Note that each lattice site is associated with two segments of the worldline, each with their own weight. To avoid double-counting,
	we follow Ref.~\cite{PhysRevD.99.074511} and by convention associate the weight of a lattice site with the \textit{outgoing} segment of the worldline
	(indicated by the arrowheads).
	}
\end{center}
\end{figure}

Accounting for all possible worldline moves, any lattice site can be represented as one of a finite set of allowed configurations.
In Fig.~\ref{Fig:Tiles}, the complete set of such configurations is depicted.
The tiles of the top two rows correspond to physical configurations; those of the bottom row exist only as intermediate
steps in the updating algorithm and are never present during a measurement of observables.
All possible transitions between these local configurations, defining the paths the worm is allowed to follow,
are collected in Table~\ref{tab:Tiles}.

\begin{table*}[]
\caption{Map of all possible local lattice updates in the worm algorithm, where integer labels correspond to those in Fig.~\ref{Fig:Tiles}.
The outgoing directions are each respective to the possible final states.}
\label{tab:Tiles}
\begin{tabular}{@{}c|c|c|c@{}}
\toprule
\multicolumn{1}{l}{\textbf{Initial state}} & \textbf{Incoming worm direction} & \textbf{Possible final states} & \textbf{Outgoing worm direction} \\ \midrule
\hline
\multicolumn{1}{c|}{\multirow{3}{*}{0}} & Up & 0, 1, 2, 3 & Down, up, left, right \\ \cmidrule(l){2-4} 
\multicolumn{1}{c|}{} & Left & 0, 5, 7 & Right, up, left \\ \cmidrule(l){2-4} 
\multicolumn{1}{c|}{} & Right & 0, 4, 6 & Left, up, right \\ \midrule
\hline
\multicolumn{1}{c|}{\multirow{3}{*}{1}} & Down & 0, 1, 2, 3 & Down, up, left, right \\ \cmidrule(l){2-4} 
\multicolumn{1}{c|}{} & Left & 5 & Down \\ \cmidrule(l){2-4} 
\multicolumn{1}{c|}{} & Right & 4 & Down \\ \midrule
\hline
\multicolumn{1}{c|}{\multirow{2}{*}{2}} & Right & 0, 1, 2, 3 & Down, up, left, right \\ \cmidrule(l){2-4} 
\multicolumn{1}{c|}{} & Left & 7 & Down \\ \midrule
\hline
\multicolumn{1}{c|}{\multirow{2}{*}{3}} & Left & 0, 1, 2, 3 & Down, up, left, right \\ \cmidrule(l){2-4} 
\multicolumn{1}{c|}{} & Right & 6 & Down \\ \midrule
\hline
\multicolumn{1}{c|}{\multirow{3}{*}{4}} & Down & 0, 4, 6 & Left, up, right \\ \cmidrule(l){2-4} 
\multicolumn{1}{c|}{} & Up & 1 & Left \\ \cmidrule(l){2-4} 
\multicolumn{1}{c|}{} & Left & 5 & Left \\ \midrule
\hline
\multicolumn{1}{c|}{\multirow{3}{*}{5}} & Down & 0, 5, 7 & Right, up, left \\ \cmidrule(l){2-4} 
\multicolumn{1}{c|}{} & Up & 1 & Right \\ \cmidrule(l){2-4} 
\multicolumn{1}{c|}{} & Right & 4 & Right \\ \midrule
\hline
\multicolumn{1}{c|}{\multirow{2}{*}{6}} & Left & 0, 4, 6 & Left, up, right \\ \cmidrule(l){2-4} 
\multicolumn{1}{c|}{} & Up & 3 & Left \\ \midrule
\hline
\multicolumn{1}{c|}{\multirow{2}{*}{7}} & Right & 0, 5, 7 & Right, up, left \\ \cmidrule(l){2-4} 
\multicolumn{1}{c|}{} & Up & 2 & Right \\ \midrule
\hline
\multicolumn{1}{c|}{\multirow{4}{*}{8}} & Up & 1 & \multirow{4}{*}{Update complete} \\ \cmidrule(lr){2-3}
\multicolumn{1}{c|}{} & Right & 4 &  \\ \cmidrule(lr){2-3}
\multicolumn{1}{c|}{} & Left & 5 &  \\ \cmidrule(lr){2-3}
\multicolumn{1}{c|}{} & Down & 0 &  \\ \midrule
\hline
\multicolumn{1}{c|}{\multirow{3}{*}{9}} & Up & 2 & \multirow{3}{*}{Update complete} \\ \cmidrule(lr){2-3}
\multicolumn{1}{c|}{} & Left & 7 &  \\ \cmidrule(lr){2-3}
\multicolumn{1}{c|}{} & Right & 0 &  \\ \midrule
\hline
\multicolumn{1}{c|}{\multirow{3}{*}{10}} & Up & 3 & \multirow{3}{*}{Update complete} \\ \cmidrule(lr){2-3}
\multicolumn{1}{c|}{} & Right & 6 &  \\ \cmidrule(lr){2-3}
\multicolumn{1}{c|}{} & Left & 0 &  \\ \bottomrule
\end{tabular}
\end{table*}

\section{Contact from the anomaly in the equation of state.}
In a $d$-dimensional non-relativistic scale-invariant system, dimensional arguments dictate that the pressure must take the form
\beq
P = \beta^{\alpha} f(\beta \mu),
\eeq
where $\alpha = -(d + 2)/ 2$. Using the thermodynamic identities
\bea
- PV &=& E - TS  - \mu N, \\
S &=& V \frac{\partial P}{\partial T}, \\
N &=& V \frac{\partial P}{\partial \mu},
\eea
where $V  = L^d$, it is straightforward to see that
\beq
- PV = E + V \beta^\alpha \alpha f(\beta \mu) = E + \alpha P V
\eeq
such that 
\beq
\label{Eq:PressureEnergyRelation}
P + 1/(\alpha + 1) E/L = 0,
\eeq
where $1/(\alpha + 1) = -2/d$. Thus, in 1D, a non-relativistic scale-invariant system obeys $P - 2 E/L = 0$.

\begin{figure}[t]
\centering
\includegraphics[width=\linewidth]{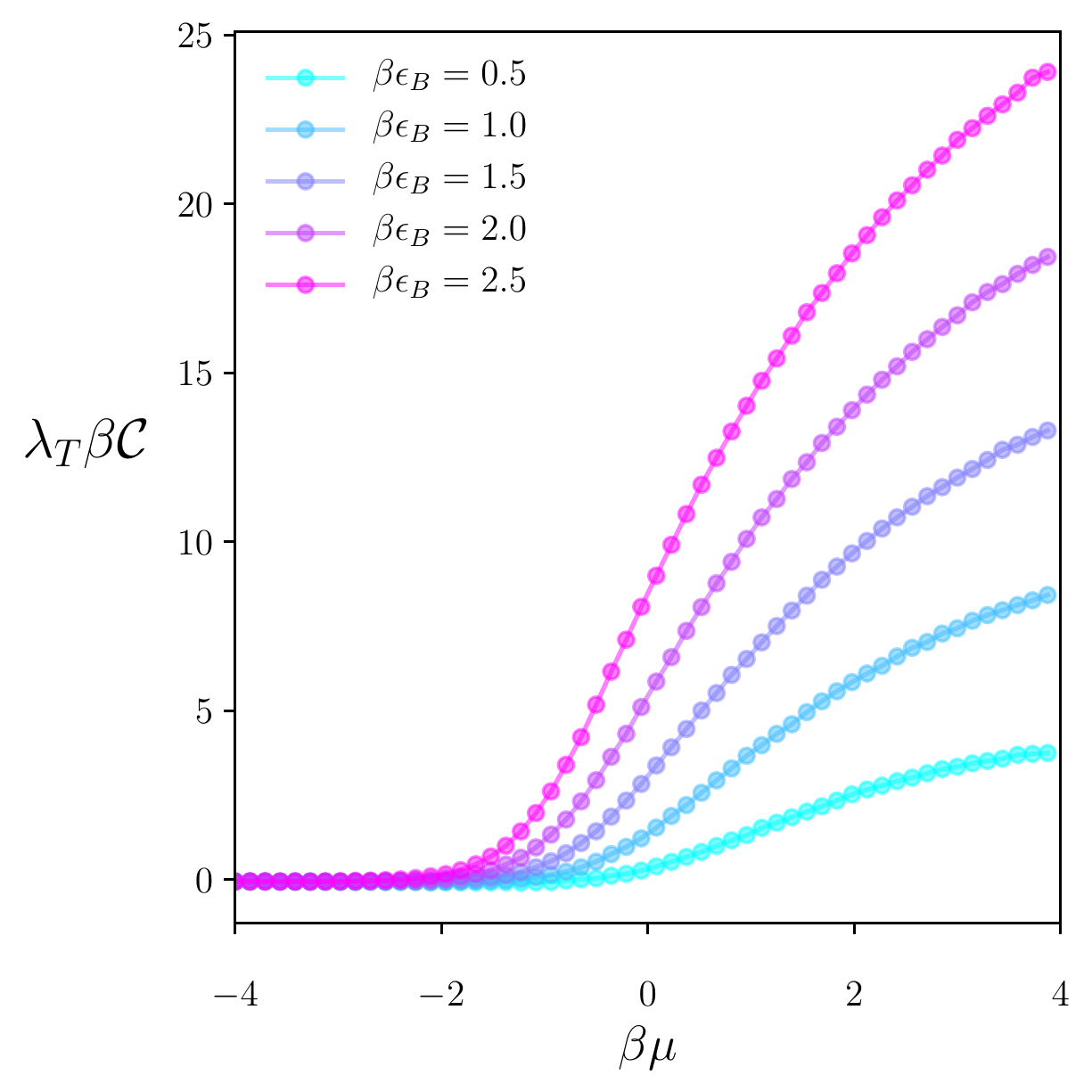}    
\caption{\label{Fig:C_diff} Contact density obtained from pressure-energy density difference.}
\end{figure}

When scale or non-relativistic invariance are broken, a non-vanishing contribution appears on the right-hand side of this
pressure-energy relation Eq.~(\ref{Eq:PressureEnergyRelation}). 
In systems with full non-relativistic invariance, that contribution is the contact density $\mathcal C$.
We have calculated that right-hand side and show it in Fig.~\ref{Fig:C_diff}. For $\beta \mu$ up to roughly $\beta \mu \simeq 0$, 
the results of Fig.~\ref{Fig:C_diff} track those of Fig.~\ref{Fig:contact}. Differences start to appear
at that point and become pronounced as $\beta \mu$ increases, likely due to lattice spacing effects which break non-relativistic
invariance. If the latter were restored, both approaches would yield the same result.

\newpage
\bibliography{WormAlg}


\end{document}